\documentclass[pre,twocolumn,10pt,aps,longbibliography,nofootinbib]{revtex4-1}

 \usepackage{verbatim}
 \usepackage{amsmath}
 \usepackage{amssymb}
 \usepackage{amsthm}
 \usepackage{latexsym}
 \usepackage{amsfonts}
 \usepackage{epsfig}
 \usepackage{epstopdf}
 \usepackage{wasysym} 
 \usepackage{color}
 \definecolor{darkblue}{rgb}{0,0,.5}
 \usepackage[linktocpage, colorlinks=true, linkcolor=darkblue, citecolor=darkblue]{hyperref}
 \usepackage[all]{hypcap}
 \usepackage[makeroom]{cancel}

\newcommand{\C}[1]{{\cal{#1}}}
\newcommand{\bb}[1]{\textbf{#1}}

\newcommand{\lr}[1]{{\left\langle {#1}\right\rangle}}

\newcommand{\id}{\operatorname{id}}

\begin{document}

\title{Stochastic thermodynamics with arbitrary interventions}

\author{Philipp Strasberg$^1$}
\author{Andreas Winter$^{1,2}$}
\affiliation{$^1$F\'isica Te\`orica: Informaci\'o i Fen\`omens Qu\`antics, Departament de F\'isica, Universitat Aut\`onoma de Barcelona, 08193 Bellaterra (Barcelona), Spain}
\affiliation{$^2$ICREA -- Instituci\'o Catalana de Recerca i Estudis Avan\c{c}ats, Pg.~Lluis Companys, 23, 08010 Barcelona, Spain}

\date{\today}

\begin{abstract}
 We extend the theory of stochastic thermodynamics in three directions: (i)~instead of a continuously monitored system 
 we consider measurements only at an arbitrary set of discrete times, (ii)~we allow for imperfect measurements 
 and incomplete information in the description, and (iii)~we treat arbitrary manipulations (e.g.~feedback control 
 operations) which are allowed to depend on the entire measurement record. For this purpose we define for a driven 
 system in contact with a single heat bath the four key thermodynamic quantities: internal energy, heat, work and 
 entropy along a single `trajectory' for a causal model. The first law at the trajectory level and the second law on 
 average is verified. We highlight the special case of Bayesian or `bare' measurements (incomplete information, but no 
 average disturbance) which allows us to compare our theory with the literature and to derive a general inequality for 
 the estimated free energy difference in Jarzynksi-type experiments. An analysis of a recent Maxwell demon 
 experiment using real-time feedback control is also given. As a mathematical tool, we prove a classical 
 version of Stinespring's dilation theorem, which might be of independent interest. 
\end{abstract}

\maketitle

\newtheorem{mydef}{Definition}[section]
\newtheorem{lemma}{Lemma}[section]
\newtheorem{thm}{Theorem}[section]
\newtheorem{crllr}{Corollary}[section]
\newtheorem*{thm*}{Theorem}
\theoremstyle{remark}
\newtheorem{rmrk}{Remark}[section]

\section{Introduction}

Stochastic thermodynamics has become a very successful theory to describe the 
thermodynamics of small, fluctuating systems arbitrarily far from equilibrium 
and even along a single stochastic trajectory (see 
Refs.~\cite{BustamanteLiphardtRitortPhysTod2005, SekimotoBook2010, JarzynskiAnnuRevCondMat2011, SeifertRPP2012, VandenBroeckEspositoPhysA2015, CilibertoPRX2017} 
for introductions and reviews). Its theoretical foundation rests on 
three pillars: (i)~the system under study is \emph{continuously} monitored, 
i.e.~the time in between two observations is effectively zero;
(ii)~the system is \emph{perfectly} measured, i.e.~there is no uncertainty left in its 
state along a single trajectory;
(iii) the system is only \emph{passively} observed, i.e.~no external 
interventions in form of disturbing measurements or feedback control 
operations are allowed. 

To the best of the authors' knowledge, no thorough study has been undertaken 
to overcome the first assumption, whereas a few interesting results have been 
obtained already to go beyond the second assumption, namely, incomplete information 
in the thermodynamic description of a stochastic system~\cite{RibezziCrivellariRitortPNAS2014, 
AlemanyRibezziCrivellariRitortNJP2015, BechhoeferNJP2015, GarciaGarciaLahiriLacostePRE2016, 
WaechtlerStrasbergBrandesNJP2016, PolettiniEspositoPRL2017, PolettiniEspositoJSP2019}. 
Beyond doubt, most progress has been achieved to incorporate feedback control, 
Maxwell's demon and different sorts of information processing 
in the description (see Refs.~\cite{ParrondoHorowitzSagawaNatPhys2015, WolpertJPA2019} for an introduction). 
Nevertheless, many feedback scenarios, such as real-time or time-delayed feedback, 
are not covered by that framework.

Here, we will show how to overcome all three assumptions (i) -- (iii) by following a recent 
proposal to build a consistent thermodynamic interpretation for a quantum stochastic 
process~\cite{StrasbergArXiv2018}. More precisely, for a system, which is possibly 
driven by an external time-dependent force and in contact with a single heat bath, 
we will equip a causal model with a consistent thermodynamic framework. 
Causal models extend the standard notion of stochastic processes where an 
external agent (e.g.~the experimenter) is not only passively observing a system, 
but where she is also allowed, e.g.~to actively intervene in the process. This 
allows for a much richer theory where correlation and causation can be 
distinguished~\cite{PearlBook2009}. We suggest to call our novel framework 
\emph{operational stochastic thermodynamics} to emphasize the fact that, from 
the perspective of the external agent, the \emph{control operations} performed 
on the system are the primary objects of interest. Here, the notion `control operation' 
is used in a wide sense and includes measurements, state preparations and feedback 
control operations, among other things. 
Only three rather standard assumptions are used here: first, in absence of any 
interventions or observations the system obeys the usual laws of thermodynamics; 
second, the system dynamics is Markovian; and third, the control operations of the 
external agent are idealized to happen instantaneously.

We note that first steps to combine stochastic thermodynamics and causal models 
have been already undertaken by Ito and Sagawa~\cite{ItoSagawaPRL2013, ItoSagawaBook2016}. 
There, stochastic thermodynamics was established for Bayesian networks, which 
are a particular representation of a causal model (here we will use a different 
one). A detailed comparison with their framework is postponed to later on. 

\emph{Outline:} As our framework requires to extend the usual notion of 
stochastic processes, Section~\ref{sec math preliminaries} gives a brief 
self-contained introduction to the mathematical theory needed in the 
following including a classical version of Stinespring's theorem. 
Section~\ref{sec op stoch thermo} then establishes the thermodynamic description 
of a causal model along a single trajectory and on average. While our theory 
is very general, it also appears quite abstract. Therefore, the special case of 
non-disturbing measurements, which is conventionally studied in the literature, 
is considered in Section~\ref{sec bare meas}. In there, we will show that 
our abstract framework allows us to draw physically relevant conclusions about, 
e.g.~the estimated free energy differences in Jarzynski-type 
experiments~\cite{LiphardtEtAlScience2002, DouarcheEtAlEPL2005, 
GuptaEtAlNatPhys2011, RamanEtAlNC2014} or the second law in a 
recently realized ``continuous Maxwell demon'' experiment~\cite{RibezziCrivellariRetortNP2019}.
Finally, we conclude with some remarks in Section~\ref{sec final remarks}. 

\section{Mathematical preliminaries}
\label{sec math preliminaries}

In classical physics, it is customary to assume that a system can be perfectly 
observed without disturbing it. If we label the elementary states of a physical 
system by $x$, then -- by measuring the system at an arbitrary set of discrete 
times $t_n>\dots>t_1$ -- we can infer the joint probability distribution 
$P(x_n,\dots,x_1)$ of finding the system in state $x_n,\dots,x_1$ at the 
respective times $t_n,\dots,t_1$. 
The assumption of a non-disturbing measurement implies the consistency condition 
\begin{equation}
  \label{eq consistency condition}
  \sum_{x_k} P(x_n,\dots,x_k,\dots,x_1) = P(x_n,\dots,\cancel{x_k},\dots,x_1),
\end{equation}
where the joint probability on the right hand side is obtained by measuring 
the system only at the times $t_n>\dots>t_{k+1}>t_{k-1}>\dots>t_1$ (i.e.~there 
is no measurement at time $t_k$). Based on this consistency condition, the 
Daniell-Kolmogorov extension theorem guarantees that there is an underlying 
continuous time stochastic process which generates the joint probabilities 
$P(x_n,\dots,x_1)$ as its marginals. 
This is the foundational cornerstone for the theory of stochastic processes, 
which allows to bridge the discrepancy between experimental reality 
(where always only a finite amount of measurement data is available) and 
its theoretical description (where the mathematical description is usually 
provided in form of a differential equation, say a master equation). 

In reality, however, an experimenter usually also influences a physical system. 
This can happen actively for a number of different reasons, e.g.~to manipulate 
a system via feedback control, to prepare a certain state of the system, or 
to learn something about the process by unravelling its causal structure 
(for instance, to test the effect of a certain drug in a clinical trial one 
usually splits the patients into two groups: those who receive the drug and 
those who receive only placebos). The experimenter can also inactively influence 
a physical system, for instance, when the measurement adds an unwanted amount 
of noise to the system, which does not vanish on average. All those examples 
violate the consistency condition~(\ref{eq consistency condition}). 

In Section~\ref{sec ctrl op}, we will review how to describe an arbitrary 
intervention or control operation performed at a single time. A causal model 
can then be seen as a set of control operations applied at different times 
to the system as reviewed in Section~\ref{sec process tensor}. Finally, for 
our thermodynamic analysis it will be important that each control operation 
can be represented in terms of more primitive operations in a larger space. 
Quantum mechanically, this representation is provided by Stinespring's theorem
and in Section~\ref{sec Stinespring} we will provide a classical analogue of it.

\subsection{Control operations}
\label{sec ctrl op}

As emphazised above, we view the terminology control operation in a broad 
sense, as any possible allowed state transformation applied to a physical system. 
The only requirement is that each control operation respects the statistical 
interpretation of the theory. 

Before we come to the most general case, it is convenient to review Bayes' theorem, 
which describes the limiting case of a `bare measurement'. By this we mean a 
measurement which is non-disturbing on average but not necessarily perfect. 
Let $p_x(t_n^-)$ be the probability to find the system in state $x\in X$ (we 
consider for definiteness only a finite state space $X$) 
prior to the measurement at time $t_n$ (in general, by $t_n^\pm$ we will denote 
the time just before or after time $t_n$). Furthermore, let $P(r_n|x)$ be the 
conditional probability to obtain result $r_n$ in the measurement given that 
the system is in state $x$. The conditional state of the system after the 
measurement is then determined by Bayes' rule, 
\begin{equation}
  \label{eq Bayes rule}
  p_x(t_n^+,r_n) = \frac{P(r_n|x)p_x(t_n^-)}{P(r_n)},
\end{equation}
where the normalization factor $P(r_n) = \sum_x P(r_n|x)p_x(t_n^-)$ denotes the
probability to obtain result $r_n$. In passing we note the slightly 
unusual notation with $p_x(t_n^+,r_n)$ denoting the conditional state of $X$ 
given the result $r_n$ [instead of using, maybe, $p_{t_n^+}(x|r_n)$], 
which, however, turns out to be beneficial later on.
Thus, whereas our state of knowledge changes 
along a single trajectory, i.e.~$p_x(t_n^+,r_n) \neq p_x(t_n^-)$, 
it does not change on average: 
\begin{equation}\label{eq Bayes rule average}
  p_x(t_n^+) = \sum_{r_n} P(r_n) p_x(t_n^+,r_n) = p_x(t_n^-).
\end{equation}
This is the essence of a non-disturbing measurement. 

In turns out to be possible to generalize the above picture to the case 
where the classical control operation also changes the state of the system 
on average. This generalized description is indeed very close to quantum 
measurement theory~\cite{WisemanMilburnBook2010}. For this purpose it is 
convenient to introduce the notion of a non-normalized system state 
$\tilde p_x(t_n^+,r_n)$, which allows to rewrite Bayes' rule as 
\begin{equation}
  \label{eq Bayes rule non norm}
  \tilde p_x(t_n^+,r_n) = \sum_{x'} A_{x,x'}(r_n) p_{x'}(t_n^-).
\end{equation}
Here, in accordance with the notation used below, we have introduced the 
matrix $A(r_n)$ with entries $A_{x,x'}(r_n) = \delta_{x,x'} P(r_n|x)$. 
In terms of the vectors $\tilde{\bb p}(t_n^+,r_n)$ and $\bb p(t_n^-)$ with
entries $\tilde p_x(t_n^+,r_n)$ and $p_{x'}(t_n^-)$, respectively, the
above can be compactly expressed as 
$\tilde{\bb p}(t_n^+,r_n) = A(r_n)\bb p(t_n^-)$.
The only difference compared to Eq.~(\ref{eq Bayes rule}) is then the 
missing normalization factor $P(r_n)$. This, in fact, implies that 
Eq.~(\ref{eq Bayes rule non norm}) is \emph{linear} with respect to the initial 
state of the system $p_x(t_n^-,r_n)$, which turns out to be convenient 
from a mathematical as well as numerical perspective. 
Furthermore, this step is of no harm, as the normalization factor is encoded 
in the non-normalized state by noting that 
$P(r_n) = \sum_x \tilde p_x(t_n^+,r_n) =: \tau \tilde{\bb p}(t_n^+,r_n)$,
introducing the probability sum operator (`trace') $\tau$.

By generalizing this picture, every possible intervention will be described 
by a set of matrices $\{A(r_n)=[A_{x,x'}(r_n)]\}$, which we call control operations. 
Each matrix $A(r_n)$ describes the action of the experimenter based on a 
(generalized) measurement result $r_n$ according to Eq.~(\ref{eq Bayes rule non norm}). 
To preserve the positivity of the unnormalized state, every control operation 
is required to satisfy $A_{x,x'}(r_n) \ge 0$, but it does no longer need 
to be diagonal. Moreover, the average effect of the control operation is 
described by a single matrix $A \equiv \sum_{r_n} A(r_n)$, i.e.~
$\bb p(t_n^+) = A \bb p(t_n^-)$, because 
\begin{equation}
  p_x(t_n^+) = \sum_{r_n} P(r_n) p_x(t_n^+,r_n) = \sum_{x'} A_{x,x'}p_{x'}(t_n^-).
\end{equation}
To preserve the statistical interpretation of the theory, $A$ is required to 
be a stochastic matrix (meaning that each column is also normalized: 
$\sum_x A_{x,x'} = 1$ for all $x'$). 
This describes the most general state transformation at the ensemble averaged 
level. Note that, in general, $p_x(t_n^+) \neq p_x(t_n^-)$. 

Hence, to conclude, classical systems which are described by probability vectors 
can be manipulated by an arbitrary set of positive matrices $\{A(r_n)\}$ with 
the sole requirement that they sum up to a stochastic matrix $A \equiv \sum_{r_n} A(r_n)$.

\subsection{Causal models}
\label{sec process tensor}

So far we focused on a single intervention happening at a single time $t_n$. 
A causal model can be seen as a prescription how to compute the effect of 
multiple interventions happening at a discrete set of times $t_n>\dots>t_1$ on the 
system. At each step some result $r_k$, $k\in\{1,\dots,n\}$, is obtained and we 
denote the entire sequence of measurement results by $\bb r_n = (r_n,\dots,r_1)$ 
in the following. Given the outcome $r_k$, we assume that the experimenter knows 
which control operation $A(r_k)$ she has implemented at time $t_k$. Moreover, we 
allow the experimenter to change her plan of interventions depending on the 
previous results $\bb r_{k-1}$ and thus, we will write $A(r_k|\bb r_{k-1})$ for 
the chosen control operation. Obviously, if the control operations describe 
bare measurements in the sense of Bayes' rule, Eq.~(\ref{eq Bayes rule}), and 
if we do not use the measurement results to manipulate the process, we recover 
the standard notion of a stochastic process. Causal models generalize this 
picture by allowing for any mathematically admissible control operation 
$A(r_k|\bb r_{k-1})$.

To add some intrinsic time-evolution of the system to the picture, we will 
assume for simplicity and in view of our thermodynamic theory in 
Sec.~\ref{sec op stoch thermo}, that the system dynamics is Markovian. This means 
that the time-evolution in between two times $t_k$ and $t_{k+1}$ can be 
described by a transition matrix $T_{k+1,k}$, which propagates the system state 
forward in time: 
\begin{equation}
 \bb p(t_{k+1}) = T_{k+1,k}\bb p(t_k).
\end{equation}
Note that Markovianity implies that the transition matrix $T_{k+1,k}$ 
is well-defined independently on which state vector its act on. 
Mathematically, $T_{k+1,k}$ is nothing else than a stochastic matrix, but to 
emphasize its dynamical role we will call it a transition matrix. We will 
make no further assumption on $T_{k+1,k}$ here.

Now, let us denote by $\bb p(t_1^-)$ the initial state of the system (which 
can be arbitrary) prior to the first control operation. The unnormalized system 
state at time $t_n^+$ after the $n$'th control operation reads 
\begin{equation}
 \begin{split}\label{eq causal model}
  \tilde{\bb p}&(t_n^+,\bb r_n)    \\
               &= A(r_n|\bb r_{n-1})T_{n,n-1}\dots A(r_2|r_1)T_{2,1}A(r_1)\bb p(t_1^-).
 \end{split}
\end{equation}
In words, the state of the system given the measurement history $\bb r_n$ is 
obtained by acting with the first control operation on it (obtaining result 
$r_1$), then letting the system evolve in time via $T_{2,1}$ until $t_2$, 
then applying the second control operation (obtaining result $r_2$), 
etc.~until time $t_n^+$. Equivalently, we can express Eq.~(\ref{eq causal model}) 
iteratively, 
\begin{equation}
 \tilde{\bb p}(t_k^+,\bb r_k) = A(r_k|\bb r_{k-1})T_{k,k-1}\tilde{\bb p}(t_{k-1}^+,\bb r_{k-1}),
\end{equation}
with $k\in\{2,\dots,n\}$ and $\tilde{\bb p}(t_1^+,r_1) = A(r_1)\bb p(t_1^-)$. 

Finally, recall that each control operation can decrease the `trace' of the 
system state and its value after the control operation is precisely the probability 
to obtain the respective measurement result. Applied to multiple control operations 
this means that the probability $P(\bb r_n)$ to obtain the sequence of results 
$\bb r_n$ is 
\begin{equation}
 P(\bb r_n) = \tau\tilde{\bb p}(t_k^+,\bb r_k) = \sum_x \tilde{p}_x(t_k^+,\bb r_n).
\end{equation}
Hence, the normalized system state after $n$ control operations is 
$\bb p(t_n^+,\bb r_n) = \tilde{\bb p}(t_n^+,\bb r_n)/P(\bb r_n)$ and the average 
system state at time $t_n^+$ reads 
\begin{equation}
 \bb p(t_n^+) = \sum_{\bb r_n} P(\bb r_n) \bb p(t_n^+,\bb r_n).
\end{equation}
Here, we used the notational convention that an averaged quantity (with respect 
to the measurement results $\bb r_n$) is denoted by simply dropping the 
dependence on $\bb r_n$ in it [as in Eq.~(\ref{eq Bayes rule average})].

To close this section, we remark that causal models can be also represented in 
different ways~\cite{PearlBook2009} (see also Sec.~\ref{sec Ito Sagawa}) and the 
picture we have given here follows closely the description in the quantum 
case~\cite{PollockEtAlPRL2018, PollockEtAlPRA2018, MilzEtAlArXiv2017}. 
The particular and simple description~(\ref{eq causal model}) 
is a consequence of the Markov property~\cite{PollockEtAlPRL2018}. Causal models 
can, however, also be defined for arbitrary non-Markovian systems, where the 
dynamics is more complicated but the control operations $A(r_n|\bb r_{n-1})$ remain 
the essential ingredients~\cite{PollockEtAlPRA2018}. A detailed comparison with 
classical causal models and the proof of a generalized extension theorem can be 
found in Ref.~\cite{MilzEtAlArXiv2017}.

\subsection{A classical version of Stinespring's theorem}
\label{sec Stinespring}
This paper aims at providing a minimal, but consistent thermodynamic description 
for an arbitrary set of control operations. Obviously, as the control operations 
can be \emph{any possible} state transformation, the resulting framework will 
on the most general level appear quite abstract. For instance, it is \emph{a priori} not clear 
how to split the energetic changes caused by the action of some control operation 
$A(r)$ into work and heat. We will see that the following theorem helps us 
fix this issue, based only on the knowledge of $A(r)$. Moreover, it is indispensable 
for finding a valid second law during the control operation. Clearly, if additional 
knowledge about the experiment is available, telling us how the control operations 
are generated physically (knowledge which we assume not to have here), the present 
description should not necessarily be taken literally. 

Moreover, we believe that the following theorem could be also useful for other 
applications. It tells us that any stochastic dynamics always arises from a reversible 
evolution in a larger space about which we have incomplete information.
It is now commonly known as Stinespring's theorem~\cite{StinespringPAMS1955}, 
but -- to the best of our knowledge -- there is no precise corresponding classical 
statement in the literature. We stress that the theorem, despite its similarity, 
does \emph{not} automatically follow from its quantum version. 

\begin{thm}
  \label{thm representation 1}
  Every stochastic matrix $A: \mathbb{R}^d\rightarrow\mathbb{R}^d$ can be represented as 
  \begin{equation}
    \label{eq representation 1}
    A\bb p = \bigl( (\id\otimes\tau)\circ \Pi \bigr) {\bb p \otimes \bb q},
  \end{equation}
  where $\bb q \in \mathbb{R}^D$ is a probability vector with a dimension  
  $D \leq d(d^2-d+1)$ and 
  $\Pi:\mathbb{R}^d\otimes\mathbb{R}^D\rightarrow\mathbb{R}^d\otimes\mathbb{R}^D$ is 
  a permutation matrix. 
  Recall that $\tau$ is the marginal (`trace') operator. 
  In terms of the matrix elements, the above equation expands to
  \[
    A_{x,x'} = \sum_{y,y'=1}^{D} \Pi_{xy,x'y'} q_{y'}.
  \]
  Moreover, for an arbitrary decomposition of $A$ into a set of control operations,
  $A(r)$ such that $A_{x,x'}(r) \ge 0$ for all $x,x'$ and $r$, we can write 
  \begin{equation}
    \label{eq representation 2}
    A(r)\bb p = \bigl[ [\id\otimes\tau B(r)]\circ \Pi \bigl] {\bb p \otimes \bb q},
  \end{equation}
  meaning
  \[
    A_{x,x'}(r) = \sum_{y,y'=1}^{D} B_{y,y}(r) \Pi_{xy,x'y'} q_{y'},
  \]
  where $B(r)$ describes the effect of a bare measurement: all $B(r)$ are 
  diagonal, non-negative matrices, whose sum is the identity matrix. 
  The permutation matrix $\Pi$ and the probability vector $\bb q$ in 
  Eq.~(\ref{eq representation 2}) are the same as in Eq.~(\ref{eq representation 1}). 
\end{thm}

In words, any stochastic evolution of a system can be seen as arising from 
marginalizing a reversible evolution in a larger `system-ancilla' space. The 
ancilla, initially described by the probabilities $q_y$, is \emph{a priori} only 
an auxiliary system without physical meaning. Often, however, it can be connected 
to a part of the real physical environment, for instance a detector or memory 
which is used to record the outcome of a measurement. Furthermore, any `selective' 
evolution conditioned on a generalized measurement result $r$ can be modeled by 
an ideal bare (or `Bayesian') measurement acting on this ancilla state only. 
This nicely encodes the fact that an experimenter usually never observes the 
system directly, but rather infers its state by looking at a secondary object,
e.g.~a display, which in turn is not affected by the observation.

A proof of Theorem~\ref{thm representation 1} is given in the Appendix, where 
we also show that the minimum dimension $D$ of the extra space is in general strictly
smaller than $d(d^2-d+1)$. We do not know how to characterize the minimum $D$,
except as a non-trivial optimization problem. 

Finally, to complete this digression, let us compare the classical with the 
quantum version of the theorem. Quantum mechanically, instead of using a stochastic 
matrix, one describes the dissipative evolution of a system by a completely positive 
and trace-preserving (cptp) map. In the extended system-ancilla space, $P$ becomes 
a unitary matrix and the dimension $D$ can be fixed to be $d^2$ (where $d$ denotes the 
dimension of the system Hilbert space). Crucially, the initial state of the ancilla 
can be always chosen to be a pure state, in which case the minimum $D$ is the 
so-called \emph{Kraus rank} of the cptp map, which coincides with the matrix 
rank of its Choi matrix. Especially the last point is in strong 
contrast to the classical version of the theorem, where a pure ancilla state 
can never suffice.

\section{Operational stochastic thermodynamics}
\label{sec op stoch thermo}

We now turn to the physical situation we wish to study and understand thermodynamically. 
We consider systems described by a finite set of states $\{x\}$, whose dynamics is 
described by a rate master equation 
\begin{equation}
  \label{eq ME}
  \frac{{\mathrm d}}{{\mathrm d}t}p_x(t) = \sum_{x'} W_{x,x'}(\lambda_t) p_{x'}(t).
\end{equation}
Here, $W(\lambda_t)$ is a rate matrix obeying $\sum_x W_{x,x'}(\lambda_t) = 0$ and 
$W_{x,x'}(\lambda_t) \ge 0$ for all $x\neq x'$. In terms of the probability
vector $\bb p(t)$, the above can be stated compactly as 
$\frac{{\mathrm d}}{{\mathrm d}t}\bb p(t) = W(\lambda_t) \bb p(t)$.

As evidenced in the notation, $W$ is allowed to depend on an external 
control parameter $\lambda_t$, which can change in time. 
Physically speaking, such a time-dependence arises from manipulating the 
free energy landscape of the system. To each state $x$ we will thus associate 
a free energy $f_x(\lambda_t) = e_x(\lambda_t) - Ts_x(\lambda_t)$, where 
$T$ is the temperature of the surrounding heat bath and $e_x(\lambda_t)$ and 
$s_x(\lambda_t)$ are the internal energy and the intrinsic entropy of state $x$, 
respectively. The intrinsic entropy arises because $x$ is not necessarily a 
`single microstate' in a Hamiltonian sense, but could be an effective mesostate 
obtained from coarse-graining over a \emph{set} of microstates (e.g. many 
microscopic configurations of a molecule can give raise to the same conformational 
state), see also Refs.~\cite{SeifertEPJE2011, EspositoPRE2012}. Furthermore, we 
assume that the rates satisfy local detailed balance, 
\begin{equation}
  \label{eq LDB}
  \frac{W_{x,x'}(\lambda_t)}{W_{x',x}(\lambda_t)} = e^{-\beta[f_x(\lambda_t) - f_{x'}(\lambda_t)]},
\end{equation}
where $\beta = T^{-1}$ (we set $k_B\equiv1$). This condition allows us to link changes in the system 
state to entropic changes in the bath. We define the following key thermodynamic 
quantities. First, the internal energy is 
\begin{equation}\label{eq internal energy}
 U(t) \equiv  \sum_x e_x(\lambda_t) p_x(t) = \bb e(\lambda_t)\cdot\bb p(t),
\end{equation}
which we have expressed as a scalar product for later convenience. Then, the heat flux and power are 
\begin{align}
  \dot Q(t) &\equiv \bb e(\lambda_t)\cdot \frac{d\bb p(t)}{dt}, \label{eq def Q} \\
  \dot W(t) &\equiv \frac{d\bb e(\lambda_t)}{dt}\cdot \bb p(t),
\end{align}
such that the first law takes on the familiar form 
\begin{equation}\label{eq first law}
 \frac{d}{dt}U(t) = \dot Q(t) + \dot W(t).
\end{equation}
Furthermore, the system entropy reads 
\begin{equation}
 \begin{split}\label{eq system entropy}
  S_S(t) &\equiv \sum_x p_x(t)[-\ln p_x(t) + s_x(\lambda_t)] \\
         &= S_\text{Sh}\left[\{p_x(t)\}_x\right] + \bb s(\lambda_t)\cdot\bb p(t),
 \end{split}
\end{equation}
where $S_\text{Sh}\left[\{p_x(t)\}_x\right] \equiv -\sum_x p_x(t)\ln p_x(t)$ denotes 
the Shannon entropy. Then, the second law of nonequilibrium thermodynamics becomes 
\begin{equation}\label{eq second law}
 \dot\Sigma(t) = 
 \frac{d}{dt}S_S(t) - \beta Q(t) - \frac{\partial\bb s(\lambda_t)}{\partial t}\cdot\bb p(t)  \ge 0,
\end{equation}
where the reversible change in intrinsic entropy needs to be substracted as it also 
appears in the time-derivative of the system 
entropy~(\ref{eq system entropy})~\cite{SeifertEPJE2011, EspositoPRE2012}. 
Furthermore, $\dot\Sigma(t)$ denotes the entropy production rate. We emphasize that 
the present setup covers a large class of systems studied in stochastic 
thermodynamics~\cite{BustamanteLiphardtRitortPhysTod2005, SekimotoBook2010, 
JarzynskiAnnuRevCondMat2011, SeifertRPP2012, VandenBroeckEspositoPhysA2015, 
CilibertoPRX2017}. What is, however, unclear at present is how to 
incorporate multiple heat reservoirs into the description. 

The goal of the present paper can now be formulated at follows. 
We consider a system, which evolves according to a Markovian rate master 
equation as above. Furthermore, we assume it obeys the laws of thermodynamics 
as specified above at the \emph{unmeasured} level (i.e. without any sort of 
interventions). Now, we seek for a consistent set of definitions of internal 
energy, heat, work and entropy along a single trajectory for an arbitrary 
causal model as described in Section~\ref{sec math preliminaries}, 
see in particular Eq.~(\ref{eq causal model}). Note that here we take 
an explicitly observer-dependent point of view: any 
action (including measurements) must be explicitly modelled within our framework, 
no further `hidden' knowledge is used. A `single trajectory' is therefore 
\emph{defined} by the sequence of measurement results $\bb r_n$, which, in general, 
refers to a discrete set of times and can include arbitrary generalized 
measurements. In view of Eq.~(\ref{eq ME}) the transition matrices in 
Eq.~(\ref{eq causal model}) are given by 
\begin{equation}\label{eq transition matrix}
 T_{k,j} \equiv \C T_+ \exp\left[\int_{t_j}^{t_k} W(\lambda_t)dt\right]
\end{equation}
with the time-ordering operator $\C T_+$. Furthermore, albeit implicit in the notation, 
we also allow the control protocol $\lambda_t = \lambda_t(\bb r_n)$ ($t\ge t_n$) to 
depend on all previous measurement outcomes. Thus, we can treat \emph{all conceivable 
feedback scenarios} within our framework. 

We remark that the sole assumptions behind the dynamical 
description~(\ref{eq causal model}) are that the system as described by Eq.~(\ref{eq ME}) 
is Markovian (but see Refs.~\cite{StrasbergEspositoPRE2019, StrasbergTBP} for 
extensions) and that the external agent effectively implements the control operations 
instantaneously, which ensures that she has full control over them. To derive a 
consistent thermodynamic interpretation for a causal model, we will use 
Theorem~\ref{thm representation 1} and model explicitly the stream of ancilla
systems interacting sequentially with the system, similar to the repeated 
interaction framework in Refs.~\cite{StrasbergEtAlPRX2017, StrasbergArXiv2018}. 

\subsection{First law}

In between two control operations, the first law is simply given by
\begin{equation}
  \label{eq first law average}
  \frac{{\mathrm d}}{{\mathrm d}t}U(t,\bb r_{n-1}) = \dot Q(t,\bb r_{n-1}) + \dot W(t,\bb r_{n-1}).
\end{equation}
This is essentially Eq.~(\ref{eq first law}), only that here we have 
explicitly emphasized that all quantities can depend on the 
entire measurement record $\bb r_{n-1}$, either because the state at time $t_{n-1}$ or the 
control protocol $\lambda_t$ (or both) depend on it. 
The time-interval of validity, $t_{n-1} < t < t_n$, is unambiguously 
indicated by the index on the sequence $\bb r_{n-1}$. 

Through the control operation at time $t_n$, non-trivial changes may happen,
as the internal energy can jump: 
\begin{equation}\label{eq first law ctrl}
 \Delta U^\text{ctrl}(\bb r_n) 
 = \bb e(\lambda_t)\cdot \left[\frac{A(r_n|\bb r_{n-1})}{P(r_n|\bb r_{n-1})} - \id\right]\bb p(t_n^-,\bb r_{n-1}).
\end{equation}
Notice that we are careful to use the normalized system state here. Hence, 
we needed to normalize the system state after the control operation by using 
the conditional probability $P(r_n|\bb r_{n-1}) \equiv P(\bb r_n)/P(\bb r_{n-1})$. 
To attribute to each control operation a meaningful heat and work, we 
make use of Theorem~\ref{thm representation 1}. The representation~(\ref{eq representation 2}), 
where the action of an arbitrary control operation is split into a reversible, 
deterministic part (the permutation matrix $\Pi$) and an irreversible, 
non-deterministic part (the bare measurement $B$), strongly suggest to associate 
changes caused by the first part as work and changes caused by the second part as 
heat~\cite{ElouardEtAlQInf2017, StrasbergArXiv2018}. However, in contrast to 
the quantum case we have to be more careful here as we are not only changing the 
energy of the system, but also its instrinsic entropy. Therefore, 
\begin{equation}
 \begin{split}
  W&^\text{ctrl}(t_n,\bb r_{n-1}) \equiv \\ 
  & \bb f(\lambda_t)\cdot\left\{ \bigl((\id\otimes\tau)\circ\Pi\bigr) \bb p(t_n^-,\bb r_{n-1})\otimes\bb q - \bb p(t_n^-,\bb r_{n-1})\right\},
 \end{split}
\end{equation}
which describes the change in the system's free energy due to the reversible and 
deterministic permutation. Note that we suppressed in the notation that the choice of the initial ancilla state 
$\bb q = \bb q(\bb r_{n-1})$ and of the permutation matrix $\Pi = \Pi(\bb r_{n-1})$ can depend on previous measurement 
results $\bb r_{n-1}$. However, due to causality 
they can not depend on the actual outcome $r_n$ obtained at time $t_n$, 
and therefore the work does not depend on it, either. 
In fact, the work during the control operation can be computed by knowing only the state after the control 
operation averaged over the last measurement result, which is 
\begin{equation}
 \begin{split}
  \bb p(t_n^+,\bb r_{n-1}) &= \sum_{r_n} P(r_n|\bb r_{n-1})\bb p(t_n^+,\bb r_n) \\
  &= \bigl((\id\otimes\tau)\circ\Pi\bigr) \bb p(t_n^-,\bb r_{n-1})\otimes\bb q \\
  &= \sum_{r_n} A(r_n|\bb r_{n-1}) \bb p(t_n^-,\bb r_{n-1}).
 \end{split}
\end{equation}
Hence, the work is uniquely determined by the average control operation $A_n \equiv \sum_{r_n} A(r_n|\bb r_{n-1})$ 
(we again suppressed the dependence on $\bb r_{n-1}$ for notational convenience) and the system state before the control 
operation. This can be compactly expressed as: 
\begin{equation}\label{eq def W ctrl}
 W^\text{ctrl}(t_n,\bb r_{n-1}) = \bb f(\lambda_t)\cdot(A_n-\id)\bb p(t_n^-,\bb r_{n-1}),
\end{equation}
Next, the heat injected during the control operation is demanded to fulfill the first law of thermodynamics 
$Q^\text{ctrl}(t_n,\bb r_n) = \Delta U^\text{ctrl}(\bb r_n) - W^\text{ctrl}(t_n,\bb r_{n-1})$. Hence, it is 
given by 
\begin{equation}
 \begin{split}\label{eq def Q ctrl}
  Q^\text{ctrl}(t_n,\bb r_n) 
  =&~ \bb e(\lambda_t)\cdot\left[\frac{A(r_n|\bb r_{n-1})}{P(r_n|\bb r_{n-1})} - A_n\right]\bb p(t_n^-,\bb r_{n-1}) \\
  &+ T\bb s(\lambda_t)\cdot(A_n-\id)\bb p(t_n^-,\bb r_{n-1}).
 \end{split}
\end{equation}
Thus, the heat depends on the last measurement result $r_n$ at the trajectory level. Averaging over it, we confirm that \begin{equation}\label{eq heat property}
 \begin{split}
  \sum_{r_n} & P(r_n|\bb r_{n-1}) Q^\text{ctrl}(t_n,\bb r_n) \\
  &= T\bb s(\lambda_t)\cdot(A_n-\id)\bb p(t_n^-,\bb r_{n-1}).
 \end{split}
\end{equation}
Finally, we remark that we have assumed that the 
states $y$ of the ancilla are energetically neutral and thus do not contribute 
to the energy balance. This is indeed the conventional choice when considering 
Maxwell demon feedbacks, where -- as we will see -- the memory of the demon can 
be identified with the ancilla. The generalization to energetically non-neutral 
ancillas is straightforward~\cite{StrasbergArXiv2018} and will not be considered here. 

To summarize, over an interval denoted by a superscript $(n]$, which starts at 
time $t_{n-1}^+$, just after the $(n-1)$-st control operation and ends at 
time $t_n^+$, just after the $n$-th control operation, the first law at the 
trajectory level can be written as usual 
\begin{equation}
  \label{eq first law complete}
  \Delta U^{(n]}(\bb r_n) = Q^{(n]}(\bb r_n) + W^{(n]}(\bb r_{n-1}),
\end{equation}
but each term is now composed out of a part referring to the unobserved evolution 
[Eq.~(\ref{eq first law average})] and to the control operation 
[Eq.~(\ref{eq first law ctrl})], e.g. 
\begin{equation}
 W^{(n]}(\bb r_{n-1}) = \int_{t_{n-1}^+}^{t_n^-} dt\dot W(t,\bb r_{n-1})+ W^\text{ctrl}(\bb r_{n-1}).
\end{equation}
The first law over multiple time-intervals 
can be obtained by concatenating the first laws for each time-interval. 

\subsection{Stochastic entropy}

Whereas we did not need to redefine the notion of internal energy, but could 
simply apply Eq.~(\ref{eq internal energy}) with respect to the conditional state of 
the system, it turns out to be necessary to \emph{redefine} the entropy of the system, 
explicitly taking into account the external ancillas and the generated measurement 
record, too. In fact, as we allow our control protocol to depend on the entire 
measurement record, it is important to store also all available information about the 
past. Thus, let us denote by $p_{x,\bb y_n}(t,\bb r_n)$ the joint probability,
conditioned on $\bb r_n$, to find the system in state $x$ and the stream of 
ancillas in state $\bb y_n = (y_n,\dots,y_1)$, where $y_k$ denotes the state of 
the ancilla responsible for the $k$-th control operation. Then, we define 
\begin{equation}\label{eq stochastic entropy}
 \begin{split}
  S(t,\bb r_n) \equiv&~ S_\text{Sh}\left[\{p_{x,\bb y_n}(t,\bb r_n)\}_{x,\bb y_n}\right] \\
  &+ \bb s(\lambda_t)\cdot\bb p(t,\bb r_n) - \ln P(\bb r_n).
 \end{split}
\end{equation}
The entropy of the system along a particular trajectory is given by three terms.
The first terms describes nothing but the remaining uncertainty about the 
system and ancilla state, which is quantified by the Shannon entropy, as usual. 
The second term simply denotes the average intrinsic entropy conditioned on 
the measurement results. The third term describes the stochastic uncertainty left 
about the measurement outcomes $\bb r_n$. When averaged over the probability 
$P(\bb r_n)$ to obtain the results $\bb r_n$, it gives the usual Shannon entropy of 
the measurement outcomes. Two important remarks are in order: 

First, while the definition~(\ref{eq stochastic entropy}) looks quite cumbersome 
in general, in can often be significantly simplified. For instance, when the final 
measurement of the ancilla system is perfect, then the information content stored in 
all ancillas is identical to the information content of the measurement results because 
\begin{equation}\label{eq perfect measurement}
 p_{x,\bb y_n}(t,\bb r_n) = p_x(t,\bb r_n)\delta_{y_n,r_n}\dots\delta_{y_1,r_1}
\end{equation}
for $t>t_n$. If the ancillas are also prepared in a zero entropy state, 
$S_\text{Sh}(\bb q) = 0$, then Eq.~(\ref{eq stochastic entropy}) reduces for all times to 
\begin{equation}
 S(t,\bb r_n) = S_\text{Sh}[\bb p(t,\bb r_n)] + \bb s(\lambda_t)\cdot\bb p(t,\bb r_n) - \ln P(\bb r_n).
\end{equation}
Furthermore, let us consider the case in which the measurement is perfect such that 
we have complete information about the system state. Then, 
$S_\text{Sh}[\bb p(t_n^+,\bb r_n)] = 0$ and if we also set $\bb s(\lambda_t) = 0$ for 
simplicity, we obtain the important limit 
\begin{equation}\label{eq stochastic entropy limit}
 S(t_n^+,\bb r_n) = - \ln P(\bb r_n).
\end{equation}

We are now in a position to compare our definition with the conventional 
one~\cite{SeifertPRL2005}, which is $-\ln p(x_t)$ where $p(x_t)$ is the probability to 
find the system in state $x$ at time $t$ as determined by the master equation~(\ref{eq ME}). 
Undeniably, this definition has turned out to be very successful and we do not want to 
question it \emph{per se} within the traditional scope of stochastic thermodynamics. 
Nevertheless, it is conceptionally not fully satisfactory. If we are confident that 
Shannon entropy is the correct thermodynamic entropy to describe a small system in 
contact with a large bath and if we have perfect knowledge about the system state, 
then its Shannon entropy should be zero and not $-\ln p(x_t)\neq0$. Furthermore, if 
information is really physical~\cite{LandauerPhysTod1991}, then it matters whether 
we measure a system or not. Thus, when we average the standard definition $-\ln p(x_t)$, 
we neglect a large part of the entropy production, which is generated in the memory of 
the measurement apparatus. 

We therefore believe that our definition~(\ref{eq stochastic entropy}) fills an 
important conceptual gap in stochastic thermodynamics. First, it reassures us that 
Shannon entropy \emph{is} the correct thermodynamic entropy for a small system as 
considered here. Second, it tells us that within the conventional (perfect measurement) 
limit of stochastic thermodynamics, the stochastic entropy is not $-\ln p(x_t)$, 
but actually $-\ln p(r_n)$. Both terms agree numerically at time $t=t_n^+$, but the latter 
corresponds to the stochastic entropy generated in the \emph{memory}. 
This should be compared with our notion~(\ref{eq stochastic entropy limit}) in the 
perfect measurement limit, which solely differs by explicitly accounting for the 
entropy generated in the entire memory. In the following we will also refer to 
Eq.~(\ref{eq stochastic entropy}) as `stochastic entropy': it is an entropy 
defined along a single trajectory and, as we will now see, gives rise to an always 
positive entropy production when averaged.

\subsection{Second law}

The second law in the absence of any control operation follows basically from 
Eq.~(\ref{eq second law}), where all quantities can now depend on the measurement 
record $\bb r_{n-1}$ as we are working at the trajectory level. Specifically, 
\begin{equation}\label{eq second law unperturbed}
 \begin{split}
  \dot\Sigma (t,\bb r_{n-1}) 
  =&~ \frac{d}{dt}S(t,\bb r_{n-1}) - \beta Q(t,\bb r_{n-1}) \\
  &- \frac{\partial\bb s(\lambda_t)}{\partial t}\cdot\bb p(t,\bb r_{n-1}) \\
  \ge&~ 0 
 \end{split}
\end{equation}
Note, however, that here we used our definition~(\ref{eq stochastic entropy}) and not 
Eq.~(\ref{eq system entropy}), which differs by taking into account the entropy 
of the ancillas and the system-ancilla correlations. Positivity of 
Eq.~(\ref{eq second law unperturbed}) is nevertheless ensured as the transition matrices 
$T_{k,j}$ of Eq.~(\ref{eq transition matrix}) act only locally on the system. Hence, they do 
not change the entropy of the ancillas and can only decrease the system-ancilla correlations. 
Proving this statement follows identical steps as in Ref.~\cite{StrasbergArXiv2018}; compare 
also with the ``modularity cost'' of Ref.~\cite{BoydMandalCrutchfieldPRX2018}.

The more interesting part concerns the entropy production during the control 
step, which we define as
\begin{equation}
  \Sigma^\text{ctrl}(t_n,\bb r_n) \equiv \Delta S^\text{ctrl}(t_n,\bb r_n) 
                                           - \beta Q^\text{ctrl}(t_n,\bb r_n),
\end{equation}
where $\Delta S^\text{ctrl}(t_n,\bb r_n) \equiv S(t_n^+,\bb r_n) - S(t_n^-,\bb r_{n-1})$ 
denotes the change in entropy due to the control operation. Note that we assume that 
the control operations happens instantaneously such that $\lambda_t$ does not vary around 
$t=t_n$. This implies that there is no reversible change in intrinsic entropy, which could 
contribute to the entropy production. It remains to be shown that the entropy 
production is positive on average, i.e. 
\begin{equation}
  \label{eq second law ctrl}
  \sum_{r_n} P(r_n|\bb r_{n-1}) \Sigma^\text{ctrl}(t_n,\bb r_n) \ge 0.
\end{equation}
This then implies $\sum_{\bb r_n} P(\bb r_n) \Sigma^\text{ctrl}(t_n,\bb r_n) \ge 0$, too. 
To prove Eq.~(\ref{eq second law ctrl}), we first notice that due to Eq.~(\ref{eq heat property}) 
and $-\ln P(\bb r_n) + \ln P(\bb r_{n-1}) = \ln P(r_n|\bb r_{n-1})$ we have 

\begin{widetext}
\begin{equation}\begin{split}
 \sum_{r_n} P(r_n|\bb r_{n-1}) \Sigma^\text{ctrl}(t_n,\bb r_n)
  &= \sum_{r_n} P(r_n|\bb r_{n-1}) 
              S_\text{Sh}\left[\bigl\{p_{x,\bb y_{n}}(t_n^+,\bb r_n)\bigr\}_{x,\bb y_{n}}\right] \\
  &\phantom{===}
               + S_\text{Sh}\left[\bigl\{P(r_n|\bb r_{n-1})\bigr\}_{r_n}\right] 
               - S_\text{Sh}\left[\bigl\{p_{x,\bb y_n}(t_n^-,\bb r_{n-1})\bigr\}_{x,\bb y_{n}}\right].
\end{split}\end{equation}
This expression characterizes the change in informational entropy of the system, the ancilla and the 
$n$-th measurement record during the control operation. We then use Eqs.~(\ref{eq representation 1}) 
and~(\ref{eq representation 2}) to write the unnormalized state of the system and all ancillas after 
the control operation as 
\begin{equation}
  \label{eq help 2nd law ctrl}
  \tilde p_{x,\bb y_n}(t_n^+,\bb r_n) 
    = P(r_n|\bb r_{n-1}) p_{x,\bb y_n}(t_n^+,\bb r_n) 
    = \sum_{x',y'_n} B_{y_n,y_n}(r_n|\bb r_{n-1}) \Pi_{xy_n,x'y_n'}(\bb r_{n-1}) 
                                      p_{x',y_n'\bb y_{n-1}}(t_n^-,\bb r_{n-1}).
\end{equation}
\end{widetext}
Note that both the permutation matrix $\Pi$ and the bare measurement $B(r_n)$ can depend 
on $\bb r_{n-1}$ if $A(r_n|\bb r_{n-1})$ depends on it. Now recall that the Shannon entropy 
is invariant under permutations and that the bare measurement in 
Eq.~(\ref{eq help 2nd law ctrl}), when summed over the outcomes $r_n$, has no effect. Thus, 
\begin{equation}\begin{split}
  &S_\text{Sh}\left[\bigl\{p_{x,y_{\bb n}}(t_n^-,\bb r_{n-1})\bigr\}_{x,\bb y_{n}}\right] \\
  &= S_\text{Sh}\left[\left\{\sum_{x',y'_n} \Pi_{xy_n,x'y'_n} 
                         p_{x',y_n'\bb y_{n-1}}(t_n^-,\bb r_{n-1})\right\}_{\!\! x,\bb y_{n}}\right] \\
  &= S_\text{Sh}\left[\left\{\sum_{r_n} P(r_n|\bb r_{n-1}) 
                         p_{x,\bb y_n}(t_n^+,\bb r_n)\right\}_{\!\! x,\bb y_{n}}\right].
\end{split}\end{equation}
The term $P(r_n|\bb r_{n-1}) p_{x,\bb y_n}(t_n^+,\bb r_n)$ can be viewed as a joint 
probability distribution over the probability space of the system, the ancilla and 
the $n$-th measurement record. But for any bipartite probability distribution 
$\{p_{ab}\}_{ab}$ with marginal $\{p_a = \sum_b p_{ab}\}_a$, we have the inequality
$S_\text{Sh}[\{p_a\}_a] \le S_\text{Sh}[\{p_{ab}\}_{ab}]$. 
Hence, Eq.~(\ref{eq second law ctrl}) is proved by noting that 
\begin{equation}\begin{split}
 S_\text{Sh}&\left[\bigl\{p_{x,\bb y_n}(t_n^-,\bb r_{n-1})\bigr\}_{x,\bb y_{n}}\right] \\
  &\leq S_\text{Sh}\left[\bigl\{P(r_n|\bb r_{n-1}) 
                                p_{x,\bb y_n}(t_n^+,\bb r_n)\bigr\}_{x,\bb y_{n},r_n}\right] \\
  &=    S_\text{Sh}\left[\bigl\{P(r_n|\bb r_{n-1})\bigr\}_{r_n}\right]                       \\
  &\phantom{=}
        + \sum_{r_n} P(r_n|\bb r_{n-1}) 
           S_\text{Sh}\left[\bigl\{p_{x,\bb y_n}(t_n^+,\bb r_n)\bigr\}_{x,\bb y_{n}}\right].
\end{split}\end{equation}

As for the first law~(\ref{eq first law complete}), the stochastic entropy production 
during the control step and during the unperturbed evolution can now be concatenated 
to give 
\begin{equation}\label{eq stochastic EP complete}
 \begin{split}
  \Sigma^{(n]}(\bb r_n) =&~ S(t_n^+,\bb r_n) - S(t_{n-1}^+,\bb r_{n-1}) - \beta Q^{(n]}(\bb r_n) \\
  &- \int_{t_{n-1}}^{t_n} dt \frac{\partial\bb s(\lambda_t)}{\partial t}\cdot p(t,\bb r_n).
 \end{split}
\end{equation}
Thus, the stochastic entropy production has the same form as in traditional 
stochastic thermodynamics, but it now involves a redefined entropy and heat 
flow. Along a single trajectory, Eq.~(\ref{eq stochastic EP complete}) can be 
negative, but on average it is always positive.

To summarize this entire section, we have introduced definitions for internal 
energy, heat, work, entropy, and entropy production along a single trajectory 
of causal models. These quantities satisfy the minimum requirements of any 
consistent theory of non-equilibrium thermodynamics: the first law holds at the 
trajectory level and the second law, with an entropy production related to entropy 
and heat in the usual way, holds on average. 

\section{The case of bare measurements}
\label{sec bare meas}

We will now consider a subclass of problems, which can be treated within 
our framework and which is close to other approaches in the literature. 
This subclass consists of control operations which are bare measurements, 
i.e.~simply updates of our state of knowledge according to Bayes' 
rule~(\ref{eq Bayes rule}). We still allow for imprecise measurements 
happening at arbitrary discrete times, thus we still have to deal with 
incomplete information similar to the situations considered in 
Refs.~\cite{RibezziCrivellariRitortPNAS2014, AlemanyRibezziCrivellariRitortNJP2015, BechhoeferNJP2015, GarciaGarciaLahiriLacostePRE2016, WaechtlerStrasbergBrandesNJP2016, PolettiniEspositoPRL2017, 
PolettiniEspositoJSP2019}. As incomplete information 
can be handled in many very different ways, we do not 
make an attempt to compare our framework in detail 
with any of those proposed in those references.
However, it is worth emphasizing that while they all deal 
with some form of incomplete information, they do not allow
any disturbing control operations. Thus, they fall into 
the class of `bare measurements'. 

Moreover, although we only observe the system, we still allow that the control protocol 
$\lambda_t$ can change depending on the measurement record (kept implicit in the notation,
as before). Thus, here we can still incorporate the conventional feedback and Maxwell 
demon scenarios~\cite{ParrondoHorowitzSagawaNatPhys2015}, which typically rely on conditioning 
$\lambda_t$ on the last measurement outcome obtained at a fixed pre-determined time. More importantly and beyond 
the standard analysis~\cite{ParrondoHorowitzSagawaNatPhys2015}, we can also treat the 
complicated cases of real-time and time delayed feedback, where the external agent can adapt her control strategy 
during the run of an experiment and where $\lambda_t$ can depend for $t>t_n$ also on $\bb r_{n-1}$ and not only on 
$r_n$. Progress in this direction was so far only achieved for model-specific 
studies~\cite{StrasbergEtAlPRE2013, MunakataRosinbergPRL2014, RosinbergMunakataTarjusPRE2015, XiaoPRE2016, 
LoosKlappSR2019}, apart from the general framework of Ref.~\cite{ItoSagawaPRL2013, ItoSagawaBook2016} to 
which we will return below in Section~\ref{sec Ito Sagawa}. 

\subsection{Stinespring's theorem for bare measurements}

One key insight of our framework was the need to model the control operations 
in a larger system-ancilla space. Hence, we will first construct this ancilla 
space for a bare measurement according to our Theorem~\ref{thm representation 1}. 
We will see that in this case the ancilla can indeed be identified with 
the degrees of freedom of a physical memory. 

We start by constructing a perfect measurement at an arbitrary time $t$ and 
add uncertainties later on. To this end, consider a $d$-dimensional ancilla 
with initial state $q_y = \delta_{y,1}$ and the permutation matrix 
$\Pi_{xy,x'y'} = \delta_{x,x'} \delta_{y,x'+y'-1}$, where the sum $x'+y'-1$ 
is in general interpreted modulo $d$. Given an arbitrary initial system 
state $p_x(t^-)$, it is straightforward to confirm that the system-ancilla state 
after the permutation is $p_{xy}(t^+) = p_x(t^-)\delta_{y,x}$, i.e.~it is 
perfectly correlated and has maximum mutual information given the marginal state $p_x(t^-)$. Finally, by applying 
a perfect measurement described by the matrix 
$B_{y,y'}(r) = \delta_{y,y'}\delta_{y,r}$, where $r\in\{1,\dots,d\}$, the 
post-measurement state of system and ancilla -- given outcome $r$ -- 
reads $p_{xy}(t^+,r) = \delta_{x,r}\delta_{y,r}$. 

Uncertainty can now be added in various ways: the ancilla could be initialized 
wrongly, we could choose a different permutation matrix or the final readout 
could be imperfect. Here, we assume that the experimenter has complete control 
over the system-ancilla interaction and can read out the state of the ancilla 
perfectly. Thus, we consider only the case where the initial ancilla state 
contains uncertainty, i.e.~$q_y \neq \delta_{y,1}$. 

\subsection{Discussion of the first and \protect\\ second law of thermodynamics}

We start with the energetics during the measurement process. From the preceeding 
section we can straightforwardly conclude that the work invested during the 
measurement, Eq.~(\ref{eq def W ctrl}), must be $0$, as we simply copy the 
system state to the ancilla and do not change the system: 
$W^\text{ctrl}(t_n,\bb r_{n-1}) = 0$. The heat exchanged during the control 
operation can, however, fluctuate along a trajectory and take on non-zero values: 
\begin{equation}
  \label{eq Q ctrl bare}
  Q^\text{ctrl}(t_n,\bb r_n) = \bb e(\lambda_t)\cdot [\bb p(t_n^+,\bb r_n) - \bb p(t_n^-,\bb r_{n-1})].
\end{equation}
That is, we interpret the random changes in energy caused by an update of our 
state of knowledge as heat, which only vanishes on average (similar to the 
`quantum heat' in Ref.~\cite{ElouardEtAlQInf2017}). 

The terminology `heat' is 
justified at least in two limiting cases. The first case is a non-driven system, 
where any change in its internal energy is due to heat: for instance, if we 
have found a two-level system in state `0' at time $t_{n-1}$ and later at time $t_n$ 
we find it in state `1', then we know that at some time $t\in(t_{n-1},t_n)$ 
the system must have jumped from state `0' to `1' by receiving an amount of heat 
$e_1-e_0$. Combining Eq.~(\ref{eq Q ctrl bare}) together with the average heat 
exchanged in between the two measurements [obtained by integrating Eq.~(\ref{eq def Q})], 
we see that our definitions exactly reproduce this intuition. 
On the other hand, if the system is driven, it was shown in Ref.~\cite{StrasbergArXiv2018} 
that in the limit of a perfect and continuous measurement, we reproduce the 
conventional definitions of stochastic 
energetics~\cite{SekimotoBook2010, SeifertRPP2012, VandenBroeckEspositoPhysA2015}, where 
the contribution~(\ref{eq Q ctrl bare}) indeed plays an essential role and cannot 
be neglected. Obviously, when the system is driven and measured only at a finite 
set of discrete times, we are leaving the realm where we can meaningfully 
compare Eq.~(\ref{eq Q ctrl bare}) with already established results, but we 
conjecture that also under these general circumstances it is justified to use 
the terminology `heat'. At least the way the term~(\ref{eq Q ctrl bare}) appears 
in the first and second law strongly suggests it. 

Next, we look at the second law during the control operation. We start with the 
change of stochastic entropy~(\ref{eq stochastic entropy}) during the control 
operation, which becomes 
\begin{equation}\begin{split}\label{eq Delta S ctrl bare}
  \Delta S^\text{ctrl}(t_n,\bb r_n) =&~S_\text{Sh}[\bb p(t_n^+,\bb r_n)] - S_\text{Sh}[\bb p(t_n^-,\bb r_{n-1})] \\
                                    & - \ln P(r_n|\bb r_{n-1}) - S_\text{Sh}(\bb q) \\
                                    & + \bb s(\lambda_t)\cdot[\bb p(t_n^+,\bb r_n) - \bb p(t_n^-,\bb r_{n-1})].
\end{split}\end{equation}
Here, $S_\text{Sh}(\bb q) = S_\text{Sh}[\{q_y\}_y]$ denotes the entropy of the 
initial ancilla state before the measurement. We tacitly assume that we are 
always implementing the same measurement (in general, $q_y$ could depend on 
$t_n$ and $\bb r_{n-1}$). Furthermore, due to the final perfect measurement of 
the ancilla, its entropy after the control operation is zero and on average 
coincides with $S_\text{Sh}[\{P(r_n|\bb r_{n-1})\}_{r_n}]$ [i.e.~the state of 
the ancillas after the measurements is identical to the measurement record $\bb r_n$, 
compare with Eq.~(\ref{eq perfect measurement})]. 
Summing up the stochastic entropy production~(\ref{eq stochastic EP complete}) over all 
intervals and using Eq.~(\ref{eq Delta S ctrl bare}), we obtain 
\begin{equation}\begin{split}
  \Sigma^\text{tot}(\bb r_n) 
     &\equiv \sum_{\ell=0}^n \Sigma^{(\ell]}(\bb r_\ell) \\
     &=       S_\text{Sh}[\bb p(t_n^+,\bb r_n)] - S_\text{Sh}[\bb p(t_0)]   \\
     &\phantom{==}
              -\ln P(\bb r_n) - nS_\text{Sh}(\bb q) - \beta Q^\text{tot}(\bb r_n) \\ 
     &\phantom{==} + \bb s(\lambda_t)\cdot[\bb p(t_n^+,\bb r_n) - \bb p(t_0)],
\end{split}\end{equation}
where $t_0<t_1$ denotes some initial time prior to the first measurement and 
$Q^\text{tot}(\bb r_n) \equiv \sum_{\ell=0}^n Q^{(\ell]}(\bb r_n)$. 
If we combine this with the integrated first law,
\begin{equation}
  \Delta U^\text{tot}(\bb r_n) 
     \equiv \sum_{\ell=0}^n \Delta U^{(\ell]}(\bb r_\ell) 
     =      Q^\text{tot}(\bb r_n) + W^\text{tot}(\bb r_n),
\end{equation}
and introduce the non-equilibrium free energy 
\begin{equation}
  F(\bb p) \equiv \bb f(\lambda_t)\cdot\bb p - T S_\text{Sh}(\bb p)
\end{equation}
for an arbitrary distribution $\bb p$ of the system, we obtain 
\begin{equation}\begin{split}
  \Sigma^\text{tot}(\bb r_n) &= \beta W^\text{tot}(\bb r_n) - \ln P(\bb r_n) - nS_\text{Sh}(\bb q) \\
                             &\phantom{==} 
                                -\beta\{F[\bb p(t_n^+,\bb r_n)] - F[\bb p(t_0)]\}.
\end{split}\end{equation}
On average, this yields the second law 
\begin{equation}\begin{split}
  \label{eq second law bare}
  \Sigma^\text{tot} &= \beta W^\text{tot} + S_\text{Sh}\left[\{P(\bb r_n)\}_{\bb r_n}\right] 
                                          - n S_\text{Sh}(\bb q)   \\
                    &\phantom{==}
                      - \beta\left\{\sum_{\bb r_n} P(\bb r_n)F[\bb p(t_n^+,\bb r_n)] 
                                          - F[\bb p(t_0)]\right\} \ge 0,
\end{split}\end{equation}
where the missing explicit dependence on the trajectory $\bb r_n$ is used 
to denote the ensemble average, that is
$\Sigma^\text{tot} = \sum_{\bb r_n} P(\bb r_n)\Sigma^\text{tot}(\bb r_n)$
and $W^\text{tot} = \sum_{\bb r_n} P(\bb r_n)W^\text{tot}(\bb r_n)$. 

We can discuss the second law~(\ref{eq second law bare}) also in view of 
other results in the literature. First, it is expressed in terms of three 
competing terms with a transparent interpretation: the work injected 
into the system, the change of entropy of the external stream of ancillas, 
and the change in free energy of the system. 
Typically, in an experiment involving feedback control, one either tries 
to extract work or to maximize the free energy of the controlled system at 
the expense of generating information in a memory. This information generation 
is exactly captured by the term 
$S_\text{Sh}\left[\{P(\bb r_n)\}_{\bb r_n}\right] - nS_\text{Sh}(\bb q)$. 
Our general second law is therefore close to the ones derived using an 
external tape of bits as an information 
reservoir~\cite{MandalJarzynskiPNAS2012, DeffnerJarzynskiPRX2013} (see also 
Ref.~\cite{StrasbergEtAlPRX2017}), whereas the mutual 
information~\cite{ParrondoHorowitzSagawaNatPhys2015} does not seem to play any 
role (compare also with the discussion in the Section~\ref{sec Ito Sagawa}). 
Furthermore, it is in general important to take into account the entire entropy 
$S_\text{Sh}\left[\{P(\bb r_n)\}_{\bb r_n}\right]$ of the measurement results 
including correlations. 

Nevertheless, there is also an important difference: the observer-dependent 
point of view including the measurement and feedback loop is explicit in our 
construction, whereas it is, at most, implicit in 
Refs.~\cite{MandalJarzynskiPNAS2012, DeffnerJarzynskiPRX2013, StrasbergEtAlPRX2017}. 
This has mathematical consequences, as our second law involves the 
\emph{average change} of free energy of the \emph{conditional system states}, whereas 
the conventionally derived second laws involving feedback control contain the 
\emph{change} of free energy of the \emph{average system state}~\cite{ParrondoHorowitzSagawaNatPhys2015, MandalJarzynskiPNAS2012, DeffnerJarzynskiPRX2013, StrasbergEtAlPRX2017}. 
Due to the convexity of entropy, these two quantities are related by the
inequality
\begin{equation}
  \sum_{\bb r_n} P(\bb r_n)F[\bb p(t_n^+,\bb r_n)] \ge F[\bb p(t_n^+)],
\end{equation}
where $\bb p(t_n^+) = \sum_{\bb r_n} P(\bb r_n)\bb p(t_n^+,\bb r_n)$ denotes the 
average system state. Thus, the second law involving the average unconditional 
system state, 
\begin{equation}\begin{split}
  \tilde\Sigma^\text{tot} 
     &\equiv \beta W^\text{tot} + S_\text{Sh}\left[\{P(\bb r_n)\}_{\bb r_n}\right] - nS_\text{Sh}[\bb q] \\
     &\phantom{==}
               - \beta\{F[\bb p(t_n^+)] - F[\bb p(t_0)]\},
\end{split}\end{equation}
is not as stringent as our second law: 
\begin{equation}
   \tilde\Sigma^\text{tot} \ge \Sigma^\text{tot} \ge 0.
\end{equation}
This makes sense: as the external agent knows the measurement record $\bb r_n$, 
the associated thermodynamic entropy (free energy) is lower (higher). 
But if that external agent passes the ensemble of systems to a second agent 
without sharing the measurement records, the uncertainty increases and 
$\tilde\Sigma^\text{tot} \ge 0$ becomes the second law associated to an 
\emph{uninformed} agent. 

\subsection{Analysis of a ``continuous Maxwell demon''}

To demonstrate the versatility of our approach, we consider the ``continuous Maxwell 
demon'' analysed in Ref.~\cite{RibezziCrivellariRetortNP2019}. As in the standard Szilard engine we consider a single 
particle in a box of volume $V$, which we partition into two subvolumes $V_0 + V_1 = V$. However, in contrast to the 
standard Szilard-type analysis we do not measure the system only once to extract an amount of work $W_0 = -T\ln V_0/V$ 
($W_1 = -T\ln V_1/V$) if we find the particle in the volume $V_0$ ($V_1$). Instead, we repeatedly measure the location 
of the particle at fixed intervals $\tau$ and extract work, when we see a \emph{change} in the particle's position 
from one compartment to the other. This is a particular example of \emph{real-time} feedback control where the external 
agent adapts her control strategy during the run of the experiment, i.e.~in each experiment the time when we extract 
work is different. Similar feedback strategies were also proposed and analysed in 
Refs.~\cite{StrasbergEtAlPRE2013, SchallerEtAlPRB2011, AverinMottonenPekolaPRB2011, EspositoSchallerEPL2012, 
ChidaEtAlNatComm2017}.

\begin{figure}[t]
 \centering\includegraphics[width=0.38\textwidth,clip=true]{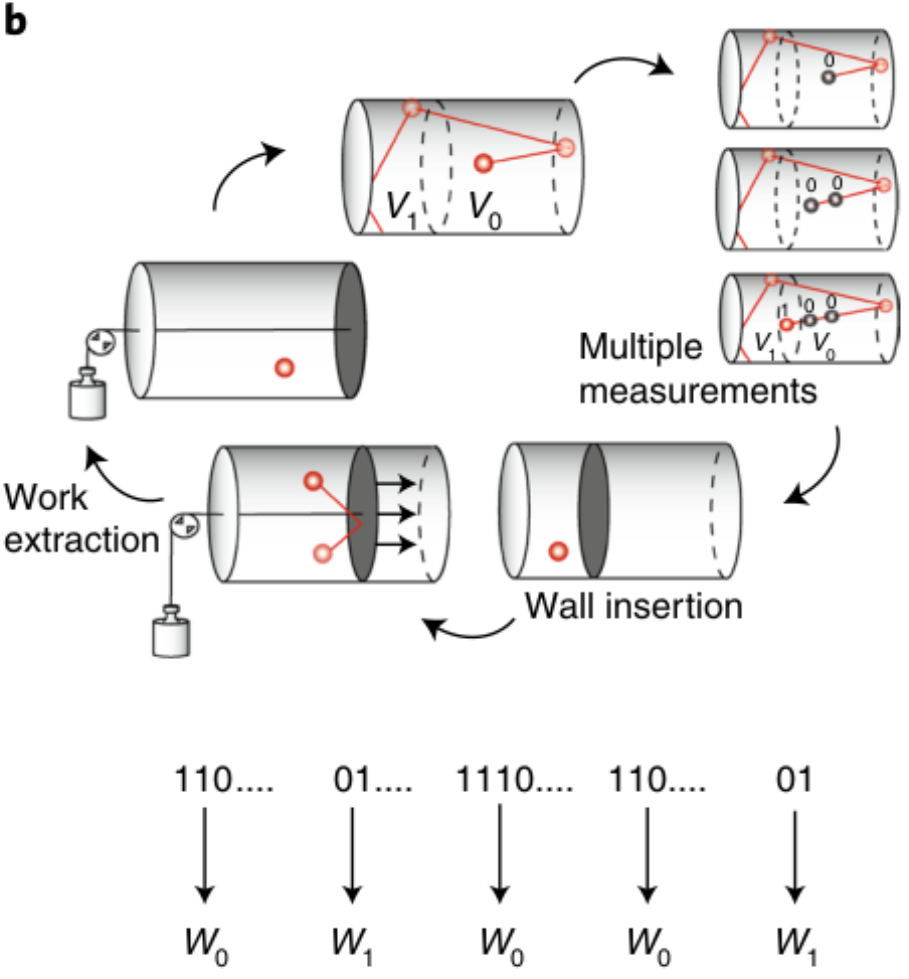}
 \label{fig CMD} 
 \caption{Sketch of the continuous Maxwell demon setup taken from 
 Ref.~\cite{RibezziCrivellariRetortNP2019} (Fig.~1b in there).}
\end{figure}

More specifically, we consider the setup shown in Fig.~\ref{fig CMD}. Initially, the particle is in equilibrium 
occupying with probability $\pi_0 = V_0/V$ ($\pi_1 = V_1/V$) the volume $V_0$ ($V_1$). Then, we perform a first 
measurement of the compartment and repeat the measurement until we see a change of the compartment. Accordingly, 
we can classify the sequence of measurement results into 0- and 1-cycles denoted by 
\begin{equation}\label{eq cycles}
 0_\ell = (\underbrace{0,\dots,0}_{\ell},1), ~~~ 1_\ell = (\underbrace{1,\dots,1}_{\ell},0),
\end{equation}
where $\ell \ge 1$ denotes the number of measurements before we measured a change of the compartment. If we have a 
0-cycle (1-cycle), we extract an amount of work $W_1$ ($W_0$) from the system such that the extracted work is on average 
\begin{equation}
 W^\text{ext} = -T(\pi_0\ln\pi_1 + \pi_1\ln\pi_0).
\end{equation}
Interestingly~\cite{RibezziCrivellariRetortNP2019}, the standard Landauer limit gives a \emph{lower} instead of an 
upper bound on the extractable work: $W^\text{ext} \ge TS_\text{Sh}[\{\pi_0,\pi_1\}]$. The resolution to this 
`paradox' is, of course, that we have to evaluate the information content of the memory with respect to the 
cycles of Eq.~(\ref{eq cycles}). In the rest of this section we are going to demonstrate that this follows 
automatically from our general framework. 

For this purpose we label by $x\in\{0,1\}$ the mesostates to find the particle in volume $V_0$ or $V_1$. 
Importantly, we have to associate an intrinsic entropy to these states given by (remember that $k_B\equiv 1$) 
\begin{equation}
 s_0 = \ln V_0/V, ~~~ s_1 = \ln V_1/V.
\end{equation}
They are computed by assuming that the single particle behaves like an ideal gas such that we can use 
$Tds = pdV$ with the pressure $p = T/V$. Furthermore, we set the entropy for the state where the particle can occupy 
the entire volume $V$ for convenience and without loss of generality to zero. Then, the extractable 
work reads $W_i = -Ts_i$ provided that the last measurement outcome is $i$. Obviously, the sequence of measurement 
results $\bb r_{\ell+1}$ is given by the cycles~(\ref{eq cycles}) with each $r_k\in\{i\} = \{0,1\}$. As in 
Ref.~\cite{RibezziCrivellariRetortNP2019} we now assume perfect measurements such that $S_\text{Sh}(\bb q) = 0$ 
where $\bb q$ denotes the initial state of the ancilla (or memory) used to record the measurement result. Furthermore, 
we do not perform any work on the system by changing some protocol $\lambda_t$, thus 
$W^\text{tot} = 0$. Note that here we equate the final extracted work $W^\text{ext}$ with the final 
change in nonequilibrium free energy when we return the system to its initial state. Therefore, $W^\text{ext}$ does 
not appear in the expression of $W^\text{tot}$. Our generalized second law~(\ref{eq second law bare}) reduces in this 
case to 
\begin{equation}
 \begin{split}
  \Sigma^\text{tot} &= S_\text{Sh}\left[\{P(\bb r_n)\}_{\bb r_n}\right] \\
                    &\phantom{==}
                      - \beta\left\{\sum_{\bb r_n} P(\bb r_n)F[\bb p(t_n^+,\bb r_n)] 
                                          - F[\bb p(t_0)]\right\} \ge 0.
 \end{split}
\end{equation}
Here, $n$ is a sufficiently large natural number in the following sense: In each run of the experiment, we will observe 
a change of the compartment at a different time $\ell\tau$, $\ell\in\mathbb{N}$. After that time we extract the work and 
restart the experiment. Now, we choose $n$ large enough such that it is almost certain that the particle has changed 
the compartment by the time $n\tau$, i.e., $P(0_n) \approx P(1_n) \approx 0$. The measurement sequences $\bb r_n$, 
e.g.~in case of a 0-cycle if the particle has changed compartment at $\ell\tau$, is then written as 
\begin{equation}
 \bb r_n(0_\ell) = (\underbrace{0,\dots,0}_{\ell},1,\underbrace{0,\dots,0}_{n-\ell-1}),
\end{equation}
where we have simply `filled in' zeros for the missing measurement results after we have aborted the experiment. 
This does not change the probability, i.e. $P[\bb r_n(0_\ell)] = P(0_\ell)$. To complete the analysis, we take into 
account that the energy of the particle does not depend on which volume it occupies such that we will conveniently set 
$e_x = 0$. Then, if we start in equilibrium as in Ref.~\cite{RibezziCrivellariRetortNP2019}, the initial free energy 
becomes 
\begin{equation}
 F[\bb p(t_0)] = -T\sum_x \pi_x(-\ln\pi_x+s_x) = 0 
\end{equation}
since $s_x = \ln V_x/V = \ln\pi_x$. On the other hand, the final free energy reads $-T\ln s_1$ ($-T\ln s_0$) in case of 
a 0-cycle (1-cycle) and appears with probability $\pi_0$ ($\pi_1$). Hence, our second law takes on the simple form 
\begin{equation}
 \begin{split}
  \Sigma^\text{tot} &= S_\text{Sh}\left[\{P(\bb r_n)\}_{\bb r_n}\right] + (\pi_0\ln\pi_1 + \pi_1\ln\pi_0) \\
  &= S_\text{Sh}\left[\{P(\bb r_n)\}_{\bb r_n}\right] - \beta W^\text{ext} \ge 0.
 \end{split}
\end{equation}
In the final step we have used that the extractable work $W^\text{ext}$ is precisely given by the change of free 
energy by letting the particle expand and thereby the system returns to its initial equilibrium configuration. 

Thus, our framework immediately leads to the desired result without the need to explicitly compute 
$S_\text{Sh}\left[\{P(\bb r_n)\}_{\bb r_n}\right]$, which was done in Ref.~\cite{RibezziCrivellariRetortNP2019} in 
order to confirm the second law. Remarkably, the above abstract experiment was realized using single molecule pulling 
experiments finding very good agreement with theory~\cite{RibezziCrivellariRetortNP2019}. 

\subsection{Fluctuation theorems}

In this paper we have so far focused on definitions for key thermodynamic quantities along a single stochastic 
trajectory, but not yet on fluctuation theorems, which are a milestone in nonequilibrium statistical 
physics~\cite{JarzynskiAnnuRevCondMat2011, SeifertRPP2012, VandenBroeckEspositoPhysA2015}. Here, we limit ourselves to 
a few observations about fluctuations in our framework.

First, the derivation of fluctuation theorems relies on a perfectly observed system state and the microreservibility of 
the underlying Hamiltonian dynamics of the system and the heat bath. For the causal model considered here, which can 
deal with any amount of uncertainty and explicitly allows to include (subjective, observer-dependent) control operations 
in the description, there is no hope of deducing a physically meaningful fluctuation theorem in general -- at least none 
which \emph{only} depends on the information available in the measurement record $\bb r_n$. Note that fluctuation 
theorems, as typically derived in the presence of feedback control~\cite{ParrondoHorowitzSagawaNatPhys2015}, still rely 
on the ability to perfectly measure the system, compare also with the discussion in 
Ref.~\cite{WaechtlerStrasbergBrandesNJP2016}. 

Second, there always exists a `formal' fluctuation theorem, which we can derive by defining a suitable `backward' or 
`time-reversed' process. For this purpose, let $\bb r_n^\dagger$ denote the sequence of measurement results $\bb r_n$ 
in reverse order and let $Q(\bb r_n^\dagger)$ be the probability to observe this sequence in the backward 
experiment, typically carried out by reversing the driving protocol $\lambda_t$. Then, given that 
$Q(\bb r_n^\dagger) = 0$ only if $P(\bb r_n) = 0$, we always have the trivial fluctuation theorem 
\begin{equation}\label{eq formal FT}
 \lr{e^{-\tilde\Sigma(\bb r_n)}}_{\bb r_n} \equiv \sum_{\bb r_n} P(\bb r_n) e^{-\tilde\Sigma(\bb r_n)} = 1,
\end{equation}
if we define the `entropy production' 
\begin{equation}
 \tilde\Sigma(\bb r_n) \equiv \ln\frac{P(\bb r_n)}{Q(\bb r_n^\dagger)}.
\end{equation}
While this quantity measures some asymmetry of the measurement statistics under time-reversal, there is no obvious 
connection of it to any thermodynamic quantity introduced above. Thus, outside the traditional limit of stochastic 
thermodynamics, Eq.~(\ref{eq formal FT}) lacks any relation to a meaningful physical quantity and therefore, does not 
share the same status as the conventional fluctuation theorem~\cite{JarzynskiAnnuRevCondMat2011, SeifertRPP2012, 
VandenBroeckEspositoPhysA2015}. 

Obviously, in the limit of a perfect and continuous bare measurement, our definitions allow to sample, e.g., the exact 
microscopic work statistics and derivations of fluctuation theorems become possible again. Remarkably, even outside this 
limit we can derive a general inequality, which links the \emph{observed} work statistics to the Jarzynski 
equality~\cite{JarzynskiPRL1997, JarzynskiPRE1997}. Let us write the observed Jarzynski equality as 
\begin{equation}
 \lr{e^{-\beta W(\bb r_n)}}_{\bb r_n} \equiv e^{-\beta\Delta F_\text{est}},
\end{equation}
where $\Delta F_\text{est}$ denotes the estimated free energy difference based on the available work statistics. 
Furthermore, let us denote by $\boldsymbol\gamma$ a system trajectory obtained from a 
perfect continuous measurement such that~\cite{JarzynskiPRL1997, JarzynskiPRE1997} 
\begin{equation}
 \lr{e^{-\beta W(\boldsymbol\gamma)}}_{\boldsymbol\gamma} = e^{-\beta\Delta F}.
\end{equation}
Finally, we introduce the conditional probability $P(\boldsymbol\gamma|\bb r_n)$ that the microscopic trajectory 
was $\boldsymbol\gamma$ given that we obtained the measurement record $\bb r_n$. Now, for bare measurements we can 
always view the observed work $W(\bb r_n)$ as resulting from a coarse-grained measurement of the perfectly measured work 
$W(\boldsymbol\gamma)$, i.e. 
\begin{equation}
 W(\bb r_n) = \sum_{\boldsymbol\gamma} W(\boldsymbol\gamma) P(\boldsymbol\gamma|\bb r_n).
\end{equation}
Since the exponential function is convex, we immediately obtain, by Jensen's inequality, 
\begin{equation}
 \lr{e^{-\beta W(\bb r_n)}}_{\bb r_n} \le \lr{e^{-\beta W(\boldsymbol\gamma)}}_{\boldsymbol\gamma},
\end{equation}
or, equivalently for the free energy differences, 
\begin{equation}
 \Delta F_\text{est} \ge \Delta F.
\end{equation}
Hence, as any experiment involves measurement errors and since the exponential function is actually strictly convex, 
we can conclude that the estimated free energy difference in any Jarzynski-type experiment \emph{always overestimates} 
the actual free energy difference: $\Delta F_\text{est} > \Delta F$. Particular estimates for the difference 
$\Delta F_\text{est} - \Delta F$ are hard to compute in general, but were worked out for particular models in 
Refs.~\cite{GarciaGarciaLahiriLacostePRE2016, WaechtlerStrasbergBrandesNJP2016}. 

\subsection{{Comparison with the framework \protect\\ of Ito \& Sagawa}}
\label{sec Ito Sagawa}

Stochastic thermodynamics of a causal model was already studied by Ito and Sagawa 
for so-called Bayesian networks~\cite{ItoSagawaPRL2013, 
ItoSagawaBook2016}. For an early study in that direction connecting 
information theory, entropy and causal models on an average level see also 
Ref.~\cite{TouchetteLloydPhysA2004}. 
Here we will outline how to connect our description to a Bayesian network and 
we will briefly highlight a few key differences in the thermodynamic description. 
A thorough comparison, however, is beyond the scope of the present paper, as in 
its most general form both frameworks, the present one and the one of 
Refs.~\cite{ItoSagawaPRL2013, ItoSagawaBook2016}, are quite involved.

Bayesian networks are a graphical representation of a probabilistic model, 
in which all random variables are specified by the nodes of the network and 
the conditional dependencies are represented by directed edges. Mathematically, 
a Bayesian network is thus given by a directed acyclic graph, which reflects 
the causal structure of the problem. Physically speaking, a directed acyclic 
graph corresponds to the fact that time `flows' in one direction and no 
time-loops are possible. The Bayesian network is fully specified once the 
probability distribution of the input variables and the conditional probabilities 
for all edges are known. 

\begin{figure}[t]
 \centering\includegraphics[width=0.42\textwidth,clip=true]{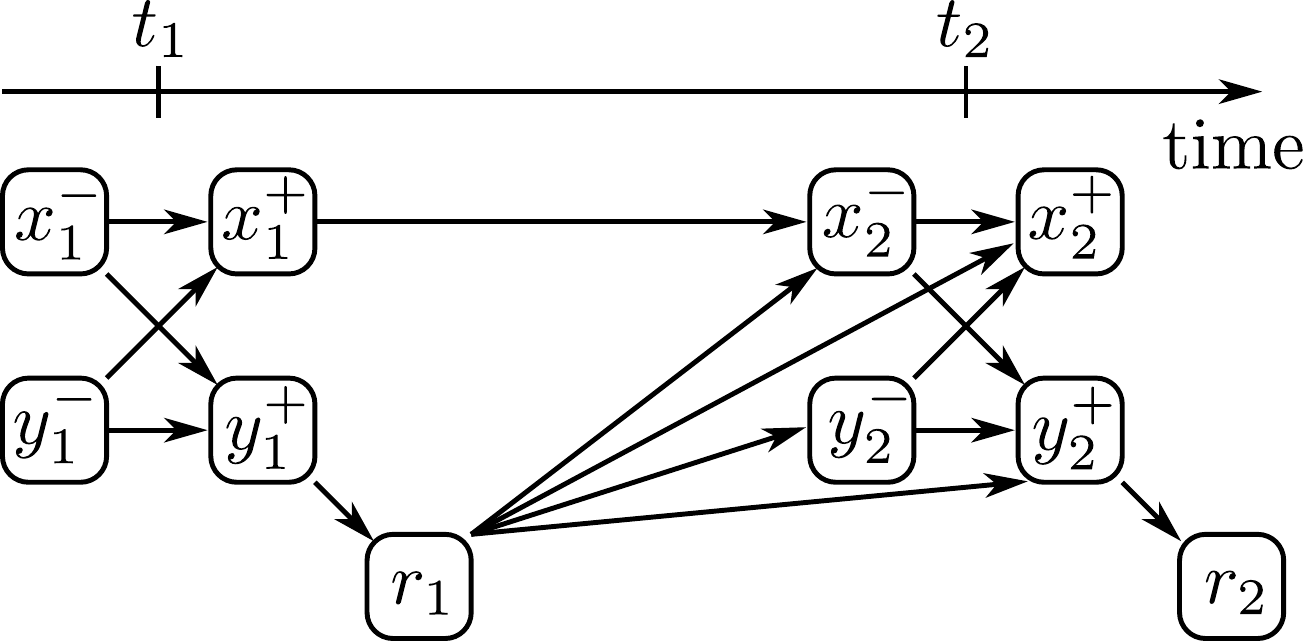}
 \label{fig Bayes network} 
 \caption{Bayesian network associated to a process with two control operations.}
\end{figure}

For illustration, we depict the Bayesian network for two control operations 
in Fig.~\ref{fig Bayes network}. It basically consists of three `layers'. 
The first layer describes the evolution of the system $X$, where we used 
$x_i^\pm$ to denote the state of the system at time $t_i^\pm$. To construct 
the control operations at time $t_i$, we use a second layer of auxiliary systems 
$Y_i$ with states $y_i^\pm$ (the ancillas). The final layer of observations by
the external agent is described by the measurement results $r_i$. 
Based on these measurement results, the external agent can decide to change 
the system evolution controlled by the protocol $\lambda_t$ or to change the 
next control operation, or both. For simplicity, we depicted only two control 
operations in Fig.~\ref{fig Bayes network} because the density of arrows in 
the picture quickly becomes very large as all previous results are allowed to 
influence the future evolution and control operations. Thus, our framework can 
be formulated in terms of a Bayesian network and could be analysed using the tools 
of Refs.~\cite{ItoSagawaPRL2013, ItoSagawaBook2016}, but there are also some essential 
differences in our setting and thermodynamic analysis. 

First, Ito and Sagawa assume that the control protocol $\lambda_t$ is constant 
in between two control operations and changes only in a step-wise fashion at 
the times $t_i$. This seems to be an essential element in their formulation in 
order to derive the second law based on the concept of a `backward trajectory', 
where transitions in the system state are required to be linked to entropy exchanges 
in the bath. Within our formalism we see that there is no need to assume that 
$\lambda_t$ remains fixed in between two control operations. 
Moreover, Ito and Sagawa assume that any change in energy due to a transition in 
the system state is due to an entropy change in the bath, see for instance Eq.~(4) 
in Ref.~\cite{ItoSagawaPRL2013}. This, however, implies that they exclude the 
possibility of any deterministic changes in the state of the system due to an external 
control operation. In other words, the work invested in the control operation is 
always zero in their case, $W^\text{ctrl}(t_n,\bb r_{n-1}) = 0$, and hence it seems 
reasonable to compare their framework with the `bare measurement' case of our framework. 

Also their thermodynamic conclusions are slightly different from ours. Apart from 
the already mentioned missing work contribution during the control step, our 
second laws are different, too. Instead of the change in entropy in the external 
stream of ancillas and the measurement record [cf. Eq.~(\ref{eq second law bare})], 
their second law contains the transfer entropy from the first layer (the system) 
to the second and third layer. The transfer entropy is an asymmetric, directed 
generalization of the mutual information concept~\cite{SchreiberPRL2000}, and therefore
the second law derived in Refs.~\cite{ItoSagawaPRL2013, ItoSagawaBook2016} is closer 
in spirit to the second law of Ref.~\cite{ParrondoHorowitzSagawaNatPhys2015}. 
In our language, their second law corresponds to the case of an `uninformed' 
agent as discussed at the end of the previous section. 

\section{Final remarks}
\label{sec final remarks}

We have provided definitions for stochastic internal energy, heat, work and entropy, 
which can be computed by an external observer who can manipulate a small system with 
arbitrary instantaneous interventions and who has no access to any further information. 
While the definition of internal energy remained the same as usual, the non-trivial effect 
of the external interventions forced us to associate a novel notion of work 
[Eq.~(\ref{eq def W ctrl})] and heat [Eq.~(\ref{eq def Q ctrl})] to it. 
Mathematically, we achieved this by using a classical version of Stinespring's 
dilation theorem (Theorem~\ref{thm representation 1}), and we ensured that we 
reproduce previous notions for already well-studied limiting cases. Hence, the 
first law at the trajectory level takes on the same form as usual and can reproduce the 
standard case of stochastic thermodynamics for a perfectly and continuously measured
system~\cite{StrasbergArXiv2018}. 

In contrast to the internal energy, we had to redefine the notion of system entropy 
from the start [Eq.~(\ref{eq stochastic entropy})]. Following the motto 
``information is physical''~\cite{LandauerPhysTod1991}, we explicitly included 
the information generated by the measurements. We then showed that the 
stochastic entropy production -- defined in the standard way as the change in 
(redefined) system entropy plus the change in entropy of the bath (which is 
proportional to the heat flow from it) -- is positive on average for any set of 
external interventions. While we do not reproduce the standard notion of stochastic 
entropy~\cite{SeifertPRL2005, SeifertRPP2012, VandenBroeckEspositoPhysA2015} in 
the respective limit, our choice guarantees that there is no need to modify the 
second law in the presence of feedback control. 

To summarize, the present paper puts forward a formally consistent framework of stochastic 
thermodynamics for an arbitrarily controlled system in contact with a single heat 
bath. Our causal model relies solely on the approximation that the external 
interventions are happening instantaneously. 
The very general, but also abstract framework of Section~\ref{sec op stoch thermo} 
allows us to draw three conclusions. First, the second law in presence of feedback control 
is more naturally expressed in terms of the Shannon entropy of the memory than the mutual 
information between the system and the memory. Second, our definition of stochastic entropy 
suggests that the stochastic entropy of Ref.~\cite{SeifertPRL2005} actually measures the 
entropy of the memory and not the system. Third, on a very abstract level, it appears that 
this framework is very similar to its quantum counterpart~\cite{StrasbergArXiv2018}, 
demonstrating that thermodynamics is a universal theory where similar principles apply to 
both, classical and quantum systems alike. In addition, we have also shown in 
Section~\ref{sec bare meas} that our theory allows to draw practically relevant conclusions.

For the future we connect the hope with our framework that it lays the foundation 
to study problems of thermodynamic inference, such as those in 
Refs.~\cite{RibezziCrivellariRitortPNAS2014, AlemanyRibezziCrivellariRitortNJP2015, 
PolettiniEspositoPRL2017, PolettiniEspositoJSP2019}, within one common unified framework. 
In that respect it would be very important to extend the present theory to multiple heat 
reservoirs too~\cite{PolettiniEspositoPRL2017, PolettiniEspositoJSP2019}. In addition, 
for practical applications it would be desirable to gain further insights into the physical 
nature of the rather abstract ancillas introduced by us.

\subsection*{Acknowledgments}

We are grateful for the useful comments of the two anonymous referees.
The authors were partially supported by the Spanish MINECO 
(project FIS2016-86681-P) with the support of FEDER funds,
and the Generalitat de Catalunya (project 2017-SGR-1127).
PS is financially supported by the DFG (project STR 1505/2-1). 


\bibliography{/home/philipp/Documents/references/books,/home/philipp/Documents/references/open_systems,/home/philipp/Documents/references/thermo,/home/philipp/Documents/references/info_thermo,/home/philipp/Documents/references/general_QM,/home/philipp/Documents/references/math_phys}

\appendix
\section{Proof of the classical \protect\\ Stinespring representation Theorem~\ref{thm representation 1}}
\label{app proof representation thm}

Our proof will be constructive and we start with the representation provided 
in Eq.~(\ref{eq representation 1}). 
For this purpose we use the fact (see Ref.~\cite{YeWangQianDCDSB2016}) that 
every stochastic matrix $A$ can be decomposed as 
\begin{equation}
  \label{eq decomposition stoch mat}
  A = \sum_{\xi=1}^N \lambda_\xi F^{(\xi)},
\end{equation} 
with $N \leq d^2-d+1$. Here, the $\lambda_\xi$ are probabilities (i.e. $\lambda_\xi\ge0$ 
and $\sum_\xi \lambda_\xi = 1$) and $F^{(\xi)}$ are deterministic transition matrices. 
This means they are binary, $F^{(\xi)}_{x,x'} \in \{0,1\}$, and they have 
exactly one `1' in each column, otherwise all entries are $0$. In general, 
$F^{(\xi)}$ is not invertible, but the set of invertible deterministic 
transition matrices coincides with the set of permutation matrices. Furthermore, 
we remark that the decomposition~(\ref{eq decomposition stoch mat}) is in general
not unique. 

To prove Eq.~(\ref{eq representation 1}), we notice that the matrix elements of 
every deterministic transition matrix can be expressed as 
$F^{(\xi)}_{x,x'} = \delta_{x,f_\xi(x')}$ where $f_\xi: X\rightarrow X$ 
is a function on the state space $X = \{1,\dots,d\}$ of the system, mapping 
$x'\in X$ to $x = f_\xi(x')\in X$, and $\delta_{x,x'}$ denotes the Kronecker delta. 
We now need to extend the set of functions $\{f_\xi\}$ to a single function, 
which is invertible and defined on a larger space $X\times Y$ where $Y$ denotes 
the state space of the ancilla. A construction that achieves this is given 
by $Y = Z\times\Xi$, where $Z = \{1,\dots,d\}$ can be regarded as a register to 
copy the state of $X$, and $\Xi = \{1,\dots,d^2-d+1\}$ labels the different functions 
used in Eq.~(\ref{eq decomposition stoch mat}). Then, we define 
\begin{equation}
 \Pi: X\times Z\times\Xi \ni (x,z,\xi) \mapsto (f_\xi(x)+z-1,x,\xi) \in X\times Z\times\Xi,
\end{equation}
where the summation in the first register is understood modulo $d$. This function 
is clearly invertible, hence it is a permutation: 
Namely, given $x$ and $\xi$, which are copied into the second and 
third register, we know $f_\xi(x)$ and from that we obtain $z$ from the 
first register. Hence, we can associate a permutation matrix $\Pi$ to it, which has elements 
\begin{equation}
 \Pi_{xz\xi,x'z'\xi'} = \delta_{x,f_{\xi'}(x')+z'-1} \delta_{z,x'} \delta_{\xi,\xi'}.
\end{equation}
Finally, we choose the initial state of the environment to be 
$q_{z\xi} = \delta_{z,1} \lambda_\xi$, which gives 
\begin{equation}
  \sum_{z,\xi,z',\xi'} \Pi_{xz\xi,x'z'\xi'} q_{z'\xi'} 
     = \sum_\xi \delta_{x,f_\xi(x')} \lambda_\xi 
     = \sum_\xi F_{x,x'}^{(\xi)} \lambda_\xi,
\end{equation}
as desired. 

Next, to prove Eq.~(\ref{eq representation 2}), we first of all note that 
any stochastic matrix $A$ can be decomposed into at most $d^2$ many different 
independent control operations $A(r)$ such that $A = \sum_r A(r)$. Any further 
control operation must then be a linear combination of the previous operations 
and as any representation of a causal model is linear in the applied operations 
$A(r)$~\cite{PollockEtAlPRA2018, MilzEtAlArXiv2017}, it 
suffices to consider $d^2$ independent ones. Thus, we choose $r$ to have two 
components, labeled $r = (\bar x,\bar x')$, and consider the elementary control 
operations $A(\bar x,\bar x')$ with elements 
\begin{equation}
  \label{eq elementary instrument}
   A_{x,x'}(\bar x,\bar x') = \delta_{x,\bar x}\delta_{x',\bar x'} A_{x,x'}.
\end{equation}
Any other decomposition of $A$ into different control operations $\tilde A(r)$ 
can be obtained from the above decomposition via linear combination, i.e. 
$\tilde A(r) = \sum_{\bar x,\bar x'} \mu_{\bar x,\bar x'}(r) A(\bar x,\bar x')$, 
for some set of positive coefficients $\{\mu_{\bar x,\bar x'}(r)\}$ which satisfy
$\sum_r \mu_{\bar x,\bar x'}(r) = 1$ for all $(\bar x,\bar x')$. 

Now we construct the bare measurement matrix $B_{y,y'}(\bar x,\bar x')$ for the 
elementary decomposition considered above. For this purpose, we introduce the subsets 
\begin{equation}
  \Xi_{x,x'} = \{\xi\in\Xi|f_\xi(x')=x\} \subset \Xi.
\end{equation}
These sets collect all $\xi\in\Xi$ which map a chosen $x'$ to a chosen $x$. We 
then define the diagonal matrix $B(r)$ via 
\begin{equation}
  B_{z\xi,z\xi}(\bar x,\bar x') 
     \equiv \sum_{\bar\xi\in\Xi_{\bar x,\bar x'}} \delta_{z,\bar x'}\delta_{\xi,\bar\xi}.
\end{equation}
Using the constructions for the permutation matrix $\Pi$ and the probability 
vector $\bb q$ from above, we confirm that 
\begin{equation}
  \label{eq help proof}
  \sum_{y,y'} B_{y,y}(\bar x,\bar x') \Pi_{xy,x'y'} q_{z'\xi'} 
     = \sum_{\bar\xi\in\Xi_{\bar x,\bar x'}} \delta_{x',\bar x'} \delta_{x,f_{\bar\xi}(\bar x')} \lambda_{\bar\xi}.
\end{equation}
On the other hand, by the definition of $A(r)$ in Eq.~(\ref{eq elementary instrument}) 
and the decomposition~(\ref{eq decomposition stoch mat}), we have 
\begin{equation}
 \begin{split}
  A_{x,x'}(\bar x,\bar x')  
    &=  \delta_{x,\bar x}\delta_{x',\bar x'} \sum_\xi \delta_{x,f_\xi(x')} \lambda_\xi   \\
    &=  \sum_{\bar\xi\in\Xi_{\bar x,\bar x'}} \delta_{x,\bar x}\delta_{x',\bar x'} 
                                              \delta_{x,f_{\bar\xi}(\bar x')} \lambda_{\bar\xi}.
 \end{split}
\end{equation}
Apart from the factor $\delta_{x,\bar x}$ this is identical to Eq.~(\ref{eq help proof}). 
But this factor is actually redundant: once we know the input $x' = \bar x'$, the output 
$x$ is fixed because the sum is restricted to only those functions which map a given 
input $\bar x'$ to a given output $\bar x$. Hence, we have proven 
Eq.~(\ref{eq representation 2}). 
\qed

\medskip
While the above construction is quite convenient, we emphasize that it is not
necessarily optimal, in the sense that in general it will be possible to find a 
representation with an ancilla space of dimension $D < d(d^2-d+1)$. 
In fact, all what we need to ensure when constructing the permutation matrix $\Pi$ 
is that, for any given output state $x\in X$ and any fixed 
decomposition~(\ref{eq decomposition stoch mat}) into $N\le d^2-d+1$ deterministic 
transition matrices, the ancilla space $Y$ has enough states to label which actual 
state $x'\in X$ was mapped to $x = f_\xi(x')$ for every possible 
$\xi\in\{1,\dots,N\}$. This would then allow us to construct an injection 
$F: X\times\{1,\dots,N\} \hookrightarrow X\times Y$, which we could extend to a bijection and 
represent by a permutation matrix $\Pi$. 
Let us denote by $|f^{-1}_\xi(x)|$ the number of elements in the preimage of $x$
under $f_\xi$. Then, the state space $Y$ must have dimension 
\begin{equation}
  D = \max_x \sum_\xi |f^{-1}_\xi(x)|,
\end{equation}
which fulfills $Nd>D\ge N$. The latter inequality implies that the state space $Y$ 
must have, for a \emph{fixed} decomposition~(\ref{eq decomposition stoch mat}), at least 
$N$ elements. To see this, consider the table of cardinalities $M_{\xi,x} \equiv |f_\xi^{-1}(x)|$. 
Because every $f_\xi$ is a function, every row of $M_{\xi,x}$ must sum up to $d$. This means 
that $Dd \ge \sum_x\sum_\xi M_{\xi,x} = \sum_\xi d = Nd$, hence $D\ge N$. 
On the other hand, the worst case scenario for a single function $\xi$ is that 
$|f_\xi^{-1}(x)| = d$, i.e. all input states get mapped to the same ouput state $x$. 
Then, $D = Nd$ implies that all functions $\xi$ map all states to the same $x$. But then 
the decomposition~(\ref{eq decomposition stoch mat}) is actually redundant as all functions 
$\xi$ were identical. Hence, we can always choose $D<Nd = d(d^2-d+1)$ implying that our 
construction above is not optimal.

Finally, we have to keep in mind that the 
decomposition~(\ref{eq decomposition stoch mat}) is not unique. Hence, the minimum 
dimension $D_\text{min}$ of the ancilla space is obtained by minimizing over all 
possible decompositions, i.e.
\begin{equation}\label{eq dimension ancilla}
  D_\text{min} = \min_{\{\xi\}} \max_x \sum_\xi |f^{-1}_\xi(x)|.
\end{equation}

Let us exemplify this reasoning in the simplest possible case of $A$ being 
a $2\times 2$ matrix, i.e. it describes a 1-bit channel. 
There are exactly four deterministic transition 
matrices: the identity, the bit-flip operation and the two matrices which map 
any input either to `0' or to `1', respectively. 
Any possible $A$ can then be written as a 
convex combination of the invertible identity map, the invertible bit-flip 
and one (but only one) of the two other non-invertible maps. 
For the invertible maps we obviously have $|f^{-1}_\xi(x)| = 1$ for every $x$ and for any of 
the non-invertible maps we have $\max_x |f^{-1}_\xi(x)| = 2$. Thus, we 
need at most $D = d^2 = 4$ ancilla states for the case of the 1-bit channel, whereas our 
explicit construction above suggested that $D = d(d^2-d+1) = 6$ is needed. 
Unfortunately, evaluating Eq.~(\ref{eq dimension ancilla}) for higher dimensions 
becomes hard very quickly.

\end{document}